# On Optimum Power Allocation for the V-BLAST

Victoria Kostina, Sergey Loyka

School of Information Technology and Engineering, University of Ottawa,

161 Louis Pasteur, Ottawa, Ontario, Canada, K1N 6N5

E-mail: sergey.loyka@ieee.org

*Abstract*— A unified analytical framework for optimum power allocation in the unordered V-BLAST algorithm and its comparative performance analysis are presented. Compact closed-form approximations for the optimum power allocation are derived, based on average total and block error rates. The choice of the criterion has little impact on the power allocation and, overall, the optimum strategy is to allocate more power to lower step transmitters and less to higher ones. High-SNR approximations for optimized average block and total error rates are given. The SNR gain of optimization is rigorously defined and studied using analytical tools, including lower and upper bounds, high and low SNR approximations. The gain is upper bounded by the number of transmitters, for any modulation format and type of fading channel. While the average optimization is less complex than the instantaneous one, its performance is almost as good at high SNR. A measure of robustness of the optimized algorithm is introduced and evaluated. The optimized algorithm is shown to be robust to perturbations in individual and total transmit powers. Based on the algorithm robustness, a pre-set power allocation is suggested as a low-complexity alternative to the other optimization strategies, which exhibits only a minor loss in performance over the practical SNR range.

*Index Terms*: MIMO system, V-BLAST, optimization, performance analysis

---

I. INTRODUCTION

The V-BLAST algorithm [1] has attracted in recent years significant attention as a signal processing strategy in the MIMO receiver due to its relative simplicity and also the ability to achieve, under certain conditions, the full MIMO capacity. Unfortunately, the algorithm has a few drawbacks as well. The optimal ordering procedure is computationally-demanding, which is a limitation for some applications. Since the successive interference cancellation is used, lower detection steps have on average a smaller SNR and thus produce more errors, which further propagate to higher steps [6], [8] so that the overall error performance may be not satisfactory, especially if no coding is used.

A popular approach to improve the error performance of the V-BLAST algorithm is to decrease the error rates at lower steps by employing a non-uniform power allocation among the transmitters[1]. Several techniques have been reported that find the transmit (Tx) power allocation that minimizes the instantaneous (i.e. for given channel realization) total error rate (TBER)[2] of the V-BLAST, with or without the optimal ordering [3]-[5]. The approximate solutions for the instantaneous Tx power allocation have also been found [3]-[5], based on various approximations. In [2], the instantaneous BLER[2] (rather than the TBER) is considered as an optimization criterion, and the optimum Tx power allocation is found numerically for the V-BLAST with two transmitters. Although the instantaneous power allocation techniques proposed in [2]-[5] do demonstrate a few dB performance improvement over the original (unoptimized) V-BLAST, they also add considerably to the system complexity, since new feedback session and power reallocation are needed each time the channel matrix changes; the instantaneous per-stream (transmitter) SNRs also need to be sent to the Tx end. A less complex approach is to use an average rather than instantaneous optimization, i.e. the optimum power allocation is found based on the average error rate (BLER or TBER). Since this ignores the small-scale fading, only occasional feedback sections and power reallocations are required, when the average SNR changes, and only the average SNR needs to be fed back to the Tx end. We adopt the latter approach in the present paper, but also study some performance measures of the instantaneous optimization for comparison purposes. We show that the average optimization provides almost the same performance improvement as the instantaneous one at high SNR, but at much smaller complexity penalty. The unordered V-BLAST is used as a baseline for optimization since (i) the optimum power allocation is considered as an alternative to the optimal ordering, and (ii) the optimal ordering presents serious difficulties for analytical performance evaluation, especially when no approximations are used [8],[16]. Similar approach has been used in [6], where the optimum power and rate allocation technique, based on minimizing the average error rate for fixed total data rate, has been proposed. However, only numerical techniques have been used there to find the optimum power and rate allocation, and also to evaluate the performance of the optimized system. On the contrary, we develop an analytical approach to the optimization of the V-BLAST power allocation[3] in the present paper, which provides more insight and is also less demanding in terms of computational power. In several cases, the developed high-SNR approximations provide reasonable accuracy in the *whole* useful range of SNR.

---

[1] A different approach to reduce the effect of error propagation has been recently proposed in [22].
[2] The TBER is defined as the error rate at the output stream to which all the individual sub-streams are merged after the detection [8]. Thus, it takes into account the actual number of errors at the transmitted symbol vector. The block error rate (BLER) is defined as the probability to have at least one error at the detected Tx symbol vector [8]. It does not take into account the *actual* number of errors, but only the fact of their presence.

Performance evaluation of the optimized systems has been done in [2]-[6] through simulations, by comparing optimized and non-optimized error rate curves, and it was noted that the optimum power allocation gives a few dB gain in terms of the SNR. We present here an analytical performance evaluation of the optimized system via a rigorous definition of the SNR gain of the optimization and via a measure of robustness, in addition to the traditional error rate analysis. It is shown that the SNR gain of the optimum power allocation cannot exceed $m$ (the number of transmitters). Additional properties of the gain, including compact high and low-SNR approximations, are also given. Since both the TBER and the BLER are widely used in the current literature as performance measures, we employ both of them and also provide a comparative analysis, which reveals many similarities and a few differences. Specifically, it is demonstrated that the TBER-based optimization results in the same performance as the BLER-based one, with the latter being more suitable for analytical techniques since it does not require explicit characterization of the error propagation effect. The impact of perturbations in the individual and total Tx powers on the performance of the optimized system is studied using a measure of robustness. It is demonstrated that the optimized system is robust to such perturbations, which also indicates that the closed-form approximations for the optimum power allocation can be used without noticeable loss in the performance. Based on this, a pre-set power allocation is suggested as a low-complexity alternative to other optimization strategies. Due to the robustness of the proposed power allocation, it is expected that a significant portion of the theoretically-predicted gain can also be achieved in practice, even when a fixed set of discrete power levels is used (as in 2G and 3G systems) instead of continuous power control. Analytical results and conclusions are validated via simulations.

---

[3] but not rate, which applies to systems with power control only and a fixed constellation. Some of our results, which are constellation-independent, can also be extended to variable-rate systems.

Overall, this paper presents a unifying analytical framework for optimization and performance analysis of the optimized unordered V-BLAST, and comparative study of various optimization strategies. The major contributions are compact closed-form approximations of the optimum power allocation, a rigorous definition and analysis of the SNR gain of optimization, and a definition and analysis of a robustness measure of the optimized algorithm.

The paper is organized as follows. Section II introduces the basic system model and gives a brief review of the relevant error rate results [6],[8], and also presents some additional expressions, which facilitate the optimization and the performance analysis. Section III states and solves the optimum power allocation problem for various optimization criteria (TBER/BLER, average/instantaneous), and studies the properties of the solutions. Section IV introduces and studies a robustness measure of the optimization algorithm. Sections V analyses the SNR gain of the optimization defined in several ways. Finally, Section VI concludes the paper.

## II. SYSTEM MODEL AND ERROR RATES

The following standard baseband discrete-time MIMO system model is employed,

$$\mathbf{r} = \mathbf{HAs} + \boldsymbol{\xi} = \sum_{i=1}^{m} \mathbf{h}_i \sqrt{\alpha_i} s_i + \boldsymbol{\xi} \qquad (1)$$

where $\mathbf{s} = [s_1, s_2, ... s_m]^T$ and $\mathbf{r} = [r_1, r_2, ... r_m]^T$ are the vectors representing the Tx and Rx symbols respectively, "T" denotes transposition, $\mathbf{H} = [\mathbf{h}_1, \mathbf{h}_2, ... \mathbf{h}_m]$ is the $n \times m$ matrix of the complex channel gains between each Tx and each Rx antenna, where $\mathbf{h}_i$ denotes i-th column of $\mathbf{H}$, $n$ and $m$ are the numbers of Rx and Tx antennas respectively, $n \geq m$, $\boldsymbol{\xi}$ is the vector of circularly-symmetric additive white Gaussian noise (AWGN), which is independent and identically distributed (i.i.d.) in each receiver[4], $\mathbf{A} = diag\left(\sqrt{\alpha_1}, ..., \sqrt{\alpha_m}\right)$, where $\alpha_i$ is the power allocated to the *i*-th transmitter. For the regular (unoptimized) V-BLAST, the total power is distributed uniformly among the transmitters, $\alpha_1 = \alpha_2 = ... = \alpha_m = 1$. In the optimized system, $\alpha_i$ are chosen to minimize the total BER or the BLER, either average or instantaneous. Since we rely on the BLAST error rate performance analysis in [6-8], we also adopt the same basic assumptions: the channel is quasistatic frequency-flat i.i.d. Rayleigh fading, the Tx signals, noise and channel gains are independent of each other; perfect channel knowledge is available at the receiver; there is no performance degradation due to synchronization and timing errors. Unless otherwise indicated, we consider BPSK modulation. Throughout the paper, we assume that the channel is a full-rank one, which is required for the V-BLAST to operate. If this is not the case, some design changes are required (e.g. changing the antenna array geometry, decreasing (increasing) the number of Tx (Rx) antennas, etc.).

---
[4] the case of unequal noise power per Rx can also be considered within the present framework, by properly re-defining $\mathbf{A}$.

The detection of a Tx symbol vector in the standard V-BLAST algorithm proceeds in steps (i.e. the *i*-th transmitter symbol is detected at step *i*) and includes 3 major procedures at each step: 1) interference cancellation from already detected symbols, 2) interference nulling from yet-to-be-detected symbols, 3) optimal ordering (based on after-detection SNR). A more detailed description of the algorithm can be found elsewhere [1]. Following [6,8], we consider the un-ordered V-BLAST in the present paper. This allows performance evaluation and optimization to be carried out analytically and in closed-form. The optimum power allocation is considered as a low-complexity alternative of the optimal ordering, which results in almost the same performance.

Analytical closed-form error performance evaluation of the un-ordered V-BLAST in uncorrelated Rayleigh fading channel have been reported in [6,8][5]. Below we outline the major results and extend them so that they can be used as a tool for the optimum power allocation. Following [8], BPSK modulation is assumed for simplicity, although the results below can also be generalized to other modulation formats (e.g M-QAM [6]), which results, however, in significantly more bulky expressions.

A. *Block Error Rate*

The BLER is defined as a probability of having at least one error in the detected Tx symbol vector, which can be expressed as [6]-[8],

$$P_B = 1 - \prod_{k=1}^{m}(1-P_{ei}) \qquad (2)$$

where $P_{ei} = P_e(\gamma_i)$ is the instantaneous, i.e. for given channel realization, conditional (no errors at the previous steps) error rate at step $i$, $\gamma_i$ is the after-processing instantaneous SNR at step $i$. The BLER is relatively easy to analyze as it does not require explicit error propagation characterization. The average (over all channel realizations) BLER $\overline{P}_B$ can be expressed in a similar way [6]-[8],

$$\overline{P}_B = 1 - \prod_{i=1}^{m}(1-\overline{P}_{ei}) \qquad (3)$$

where $\overline{P}_{ei} = \langle P_{ei} \rangle_\mathbf{H}$ is the average conditional error rate at step i, which is the same as the average error rate with ($n-m+i$)-th order maximum ratio combining (MRC), and is known in closed form for many modulation formats [18]. Specifically, for BPSK modulation,

$$\overline{P}_{ei} = \overline{P}_{(n-m+i)}^{MRC}(\alpha_i\gamma_0) = \left[\frac{1-\mu_i}{2}\right]^{n-m+i} \sum_{k=0}^{n-m+i-1} C_{n-m+i-1+k}^{k}\left[\frac{1+\mu_i}{2}\right]^{k}, \quad \mu_i = \sqrt{\frac{\alpha_i\gamma_0}{1+\alpha_i\gamma_0}}, \qquad (4)$$

where $\gamma_0 = 1/\sigma_0^2$ and $\alpha_i\gamma_0$ are the average per-Tx SNR for the unoptimized and optimized V-BLAST, and $\sigma_0^2$ is the noise variance in each receiver.

---

[5] While [6] considers the QR-based V-BLAST, it can be shown that its performance is identical to the regular (zero-forcing successive interference cancellation) V-BLAST [14, 8].

### B. Total Bit Error Rate

The TBER, i.e. the error rate in the output data stream to which all the individual Tx streams are merged, is given by

$$P_{et} = \frac{1}{m}\sum_{i=1}^{m} P_{ui}, \qquad (5)$$

where $P_{ui} = P_{ui}(\gamma_1...\gamma_i)$ is the unconditional error probability at step $i$, which includes the errors propagating from the preceding steps. To account for different combinations of errors in the first $i$-1 steps, the error vector is introduced: $\mathbf{E}_{i-1} = [e_1, e_2...e_{i-1}]$, where $e_k = \hat{s}_k - s_k$ represents demodulation error at step $k$, $e_k \in \{0, \pm 2\}$, and $\hat{s}_k$ denotes the symbol demodulated at step k. $P_{ui}$ can then be expressed as:

$$P_{ui} = \sum_{\mathbf{E}_{i-1}} P_{ei|\mathbf{E}_{i-1}} P_{\mathbf{E}_{i-1}}, \qquad (6)$$

where $P_{ei|\mathbf{E}_{i-1}} = P_{ei|\mathbf{E}_{i-1}}(\gamma_i)$ is the probability of error at $i$-th step conditioned on the error vector $\mathbf{E}_{i-1}$, and $P_{\mathbf{E}_{i-1}} = P_{\mathbf{E}_{i-1}}(\gamma_1...\gamma_{i-1})$ is the probability that such error vector occurs, which can be expressed as:

$$P_{\mathbf{E}_{i-1}} = \prod_{k=1}^{i-1} \Pr\{e_k \mid \mathbf{E}_{k-1}\}, \quad \Pr\{e_k \mid \mathbf{E}_{k-1}\} = \begin{cases} P_{ek \mid \mathbf{E}_{k-1}}, & e_k \neq 0 \\ 1 - P_{ek \mid \mathbf{E}_{k-1}}, & e_k = 0 \end{cases} \quad (7)$$

Rather than considering the TBER as a sum of unconditional BERs $P_{ui}$ as in (5) [8], where each $P_{ui}$ includes the errors propagating from steps 1 to $i-1$, the sum in (5) can be regrouped to emphasize the errors that occurred for the first time at step $i$ and then propagated further to steps $i+1,\ldots,m$. To this end, let us regroup the error vectors by the position of the first error, i.e. group $i$ includes all error vectors that can be written as $\tilde{\mathbf{E}}_{j-1} = [0,\ldots 0, \pm 2, e_{i+1} \ldots e_{j-1}]$. The TBER can then be written as

$$P_{et} = \frac{1}{m} \sum_{i=1}^{m} a_i P_{ei} \prod_{k=1}^{i-1} (1 - P_{ek}),$$

$$a_i = 1 + P_{e_{i+1} \mid \pm 2} + \sum_{j=i+2}^{m} \sum_{[e_{i+1} \ldots e_{j-1}]} P_{ej \mid \tilde{\mathbf{E}}_{j-1}} \prod_{k=i+1}^{j-1} \Pr\{e_k \mid \tilde{\mathbf{E}}_{k-1}\}, \quad a_m = 1 \quad (8)$$

where $a_i$ describes the after-effects of the error that first occurred at step $i$. If there were no error propagation, then all $a_i = 1$; the error propagation effect increases $a_i$, resulting in higher TBER. In many cases (i.e. intermediate to high SNR), (8) is easier to deal with than (5)-(6), since simple but accurate approximations are straightforward to obtain.

Note that (5)-(8) hold for both instantaneous and average error rates. In the latter case, similarly to [8], the average step BER conditioned on $\mathbf{E}_{i-1}$ is:

$$\bar{P}_{ei \mid \mathbf{E}_{i-1}} = \bar{P}_{(n-m+i)}^{MRC}\left(\gamma_i^{eff}\right), \quad \gamma_i^{eff} = \frac{\alpha_i}{\left|\mathbf{E}_{i-1}\mathbf{A}_{i-1}\right|^2 + \sigma_0^2}, \quad (9)$$

where the unequal power distribution is taken into account by using $\mathbf{A}_{i-1} = diag\left(\sqrt{\alpha_1},\ldots,\sqrt{\alpha_{i-1}}\right)$, and $\gamma_i^{eff}$ is the "effective step SNR", which includes propagating errors from the past decisions as interference. Finally, the average TBER can be evaluated using (9) in (5)-(8). If there is no error in the earlier decisions, $\gamma_i^{eff} = \alpha_i/\sigma_0^2 = \alpha_i \gamma_0$. At intermediate to large SNR, earlier errors greatly reduce $\gamma_i^{eff}$ since $\left|\mathbf{E}_{i-1}\mathbf{A}_{i-1}\right| \gg \sigma_0$.

To find the instantaneous TBER, we need to evaluate $P_{ei \mid \mathbf{E}_{i-1}}$. This can be accomplished by considering the decision variable at step $i$ (after the interference cancellation and nulling),

$$\hat{r}_i = \mathbf{w}_i^+ \mathbf{h}_i \sqrt{\alpha_i} s_i + \mathbf{w}_i^+ \sum_{j=1}^{i-1} \mathbf{h}_j \sqrt{\alpha_j} e_j + \mathbf{w}_i^+ \xi \quad (10)$$

where $^+$ denotes Hermitian conjugate, $\mathbf{w}_i$ are the optimum combining weights that completely eliminate the inter-stream interference from yet-to-be-detected symbols and maximize the output SNR [8]. For the binary decision rule applied to $\hat{r}_i$, one obtains:

$$P_{ei|\mathbf{E}_{i-1}} = \frac{1}{2}P(\text{Re}\{\hat{r}_i\} > 0 | s_i = -1, \mathbf{E}_{i-1}) + \frac{1}{2}P(\text{Re}\{\hat{r}_i\} < 0 | s_i = +1, \mathbf{E}_{i-1}) =$$
$$= \frac{1}{2}Q\left(\text{Re}\{\mathbf{w}_i^+\mathbf{h}_i\sqrt{\alpha_i} - \mathbf{w}_i^+\sum_{j=1}^{i-1}\mathbf{h}_j\sqrt{\alpha_j}e_j\}\right) + \frac{1}{2}Q\left(\text{Re}\{\mathbf{w}_i^+\mathbf{h}_i\sqrt{\alpha_i} + \mathbf{w}_i^+\sum_{j=1}^{i-1}\mathbf{h}_j\sqrt{\alpha_j}e_j\}\right), \quad (11)$$

where $Q(x) = \sqrt{2\pi}^{-1}\int_x^\infty e^{-t^2/2}dt$ is the Q-function[6]. The instantaneous TBER is obtained by substituting (11) into (5)-(8).

### C. Average Error Rates in High SNR Region

An accurate high-SNR approximation of (3) can be obtained by approximating $\overline{P}_{ei}$ and keeping only the leading term in each $\alpha_i$,

$$\overline{P}_B(\boldsymbol{\alpha}) \approx \sum_{i=1}^m \overline{P}_{ei} \approx \sum_{i=1}^m \frac{C_{2i-1}^i}{(4\alpha_i\gamma_0)^{n-m+i}}, \quad (12)$$

Due to increasing diversity order with step number in un-optimized V-BLAST ($\alpha_1 = \alpha_2 = ... = \alpha_m = 1$), the average BLER is well approximated by the first step BER [8],

$$\overline{P}_B \approx \overline{P}_{e1} \text{ for } \gamma_0 \gg 1, \quad (13)$$

As we demonstrate below, this approximation also holds for the optimized system (see section III-B).

Using (8), the high-SNR approximation of the average TBER for the un-optimized systems is as follows [8]:

$$\overline{P}_{et} \approx \frac{\overline{a}_1}{m}\overline{P}_{e1}, \quad \overline{a}_1 = 1 + \overline{P}_{e2|\pm 2} + \sum_{i=3}^m \sum_{[e_2...e_{i-1}]} \overline{P}_{ei|\tilde{\mathbf{E}}_{i-1}}\prod_{k=2}^{i-1}\Pr\{e_k | \tilde{\mathbf{E}}_{k-1}\}, \quad \overline{P}_{ei|\tilde{\mathbf{E}}_{i-1}} \approx \overline{P}_{(n-m+i)}^{MRC}\left(\left|\tilde{\mathbf{E}}_{i-1}\right|^{-2}\right), \quad (14)$$

where only the error patterns with an error at 1$^{st}$ step, $\tilde{\mathbf{E}}_{i-1} = [\pm 2, e_2...e_{i-1}]$, are considered; $\overline{a}_1$ quantifies the contribution of the error propagation effect to the TBER and is independent of the average SNR. This expression will be instrumental in comparing the impacts of error propagation on the performance of the non-optimized and optimized systems. Since the 1$^{st}$ step BER has a dominant effect on the overall performance in the high SNR

---

[6] Due to the fast-decaying behavior of the Q-function, $P_{ei|\mathbf{E}_{i-1}}$ is well approximated by

$P_{ei|E_{i-1}} \approx Q(x_{\min})/2$, $x_{\min} = \min\left[\text{Re}\{w_i^+h_i\sqrt{\alpha_i} \pm w_i^+\sum_{j=1}^{i-1}h_j\sqrt{\alpha_j}e_j\}\right]$

region, whether TBER or BLER is used as a performance criterion, it follows then that optimum power allocation algorithm should reduce the 1$^{st}$ step BER by allocating most of the power to the 1$^{st}$ transmitter. As we demonstrate below, this is indeed the case.

For the optimized system at high SNR, the average TBER can be well approximated by using the high-SNR approximations of $\overline{P}_{ei}$ (similarly to the average BLER approximation) and the approximated error propagation probability, $\overline{P}_{ei|\mathbf{E}_{i-1}} \approx 1/2$ if $|\mathbf{E}_{i-1}|^2 \neq 0$ ; see section III-C for further details.

## III. OPTIMUM POWER ALLOCATION

Under the total Tx power constraint, individual (per Tx or stream) powers can be optimally allocated in such a way as to minimize the TBER or the BLER, either instantaneous or average. While the instantaneous (i.e. for each channel realization) power allocation requires an instantaneous feedback in order to supply the Tx end with the optimum allocation for each channel realization, the average power allocation does not require instantaneous feedback (only the average SNR needs to be known at the Tx end) and hence does not incur significant penalty in complexity. Below we provide a comparative analysis of the optimizations based on the BLER and the TBER, both instantaneous and average.

The problem of optimum power allocation can be formulated as follows:

$$\text{minimize } P(\boldsymbol{\alpha}), \text{ subject to } \sum_{i=1}^{m} \alpha_i = m, \tag{15}$$

where $P$ is the objective function equal to the BLER or the TBER, either instantaneous or average, whose argument is the power allocation coefficients $\boldsymbol{\alpha} = [\alpha_1, \alpha_2, \ldots \alpha_m]^T$ [7]. For the non-optimized system (uniform power allocation), $\alpha_1 = \alpha_2 = \ldots = \alpha_m = 1$.

### A. Uniqueness of the Solution

The optimization problem in (15) is a convex one and thus has a unique (global) solution when the BLER is used as the objective, $P(\boldsymbol{\alpha}) = P_B(\boldsymbol{\alpha})$ or $P(\boldsymbol{\alpha}) = \overline{P}_B(\boldsymbol{\alpha})$. The uniqueness of the solution for $P(\boldsymbol{\alpha}) = \overline{P}_B(\boldsymbol{\alpha})$ has been demonstrated in [6][8]. In the case of the instantaneous BLER, $P(\boldsymbol{\alpha}) = P_B(\boldsymbol{\alpha})$, the uniqueness follows from the same argument since $P_{ei} = P_e(\gamma_i)$ is a convex function of $\gamma_i$ for BPSK modulation. This significantly simplifies the analysis since any found solution is automatically the global minimum.

---

[7] It is also a function of $\gamma_0$ and, in the case of instantaneous optimization, of the instantaneous per-stream SNRs

[8] while QAM modulation has been considered in [6], the argument there holds true for BPSK modulation as well, by noting that $Q(\sqrt{\gamma})$ is a convex function. We note that this result also extends, in a non-trivial way, to any 1-D or 2-D

If the instantaneous TBER is used as an objective, it can be shown through numerical evidence that it can be non-convex, depending on a channel realization, i.e. some channel realizations produce convex $P(\boldsymbol{\alpha}) = P_{et}(\boldsymbol{\alpha})$, and some – non-convex, so that the uniqueness of the solution cannot be guaranteed. While the problem can be solved numerically anyway, care should be taken when choosing a starting point in the numerical algorithm as it affects to which particular local minimum the algorithm will converge. It may also affect the convergence speed of the algorithm. To ensure that the numerical algorithm employed finds the global optimum in most cases, we use in our numerical simulations below several starting points, which results in several local minima, and choose the best one.

In the case of average TBER, the uniqueness of the solution is an open problem. Extensive numerical evidence indicates that it is unique (i.e. only one global minimum). At high average SNR, the uniqueness follows from the high-SNR approximation of the TBER, which is convex (see subsection C for details).

### B. Power Allocation Optimum in the Average BLER

In this case, the average BLER is the objective function in (15), $P(\boldsymbol{\alpha}) = \overline{P}_B(\boldsymbol{\alpha})$. Using the Lagrange multiplier technique for constrained optimization with the following Lagrangian[9],

$$L(\boldsymbol{\alpha}) = \overline{P}_B(\boldsymbol{\alpha}) + \lambda \left( \sum_{i=1}^{m} \alpha_i - m \right), \quad (16)$$

the optimum $\boldsymbol{\alpha}$ are found from

$$\partial L(\boldsymbol{\alpha})/\partial \alpha_i = \partial \overline{P}_B(\boldsymbol{\alpha})/\partial \alpha_i + \lambda = 0, \ i = 1...m \quad (17)$$

where $\lambda \geq 0$ is the Lagrange multiplier, which is found from the total power constraint, $\sum_{i=1}^{m} \alpha_i(\lambda) = m$, i.e. (17) and the constraint are considered together as a system of equations[10]. Since the objective is convex, the solution is the global minimum. Using (3) and after some manipulations, (17) reduces to,

---

constellation, and also to some multi-dimensional constellations [15].
[9] similar approach to optimization of power allocation has been developed in [19].
[10] strictly speaking, an additional constraint $\alpha_i \geq 0$, $i = 1...m$, is required. However, since our solutions always satisfy it, we do not include it explicitly.

$$\frac{\partial \ln(1-\overline{P}_{ei})}{\partial \alpha_i} = -\frac{1}{1-\overline{P}_{ei}} \frac{\partial \overline{P}_{ei}}{\partial \alpha_i} = \frac{\lambda}{1-\overline{P}_B}, \ i=1...m \tag{18}$$

Note that the optimality conditions (18) do not require all $\overline{P}_{ei}$ be equal; rather, their normalized derivatives should be equal. Unfortunately, (18) together with the total power constraint is a system of nonlinear transcendental equations, which in general cannot be solved analytically in closed form. Thus, some approximations are required.

For high SNR, (18) can be approximated as

$$\frac{\partial \ln(1-\overline{P}_{ei})}{\partial \alpha_i} \approx -\frac{\partial \overline{P}_{ei}}{\partial \alpha_i} \approx \lambda, \ i=1...m \tag{19}$$

so that the optimality requires all the derivatives $\partial \overline{P}_{ei}/\partial \alpha_i$ be equal, which is what one would intuitively expect based on the total power constraint.

A compact and accurate analytical solution of (19) can be obtained using the Newton-Raphson method and the approximation in (12) (see Appendix A for details):

$$\alpha_i^{opt} \approx \frac{m\tilde{\alpha}_i}{\sum_{k=1}^{m}\tilde{\alpha}_k}, \ \tilde{\alpha}_i \approx \frac{b_i}{(4\gamma_0)^{\frac{i-1}{n-m+i+1}}}\left(1 - \frac{b_2}{mc_1^{n-m+\sqrt[3]{4\gamma_0}}}\right)^{c_i}, \tag{20}$$

where the numerical coefficients $b_i$, $c_i$ are given by

$$b_i = \sqrt[n-m+i+1]{\frac{(n-m+i)m^{n-m+2}C_{2i-1}^i}{n-m+1}}, \ c_i = \frac{(n+1)!}{(n-m+1)!(n-m+i+1)} \tag{21}$$

The fist equality in (20) assures the total power constraint holds for the approximated allocation. Noting that $b_1 = m$, (20) can be further approximates as,

$$\alpha_1^{opt} \approx m - \sum_{i=2}^{m}\alpha_i^{opt}, \ \alpha_i^{opt} \approx \frac{b_i}{(4\gamma_0)^{\frac{i-1}{n-m+i+1}}}, \ i=2,...,m, \tag{22}$$

i.e. almost all the power goes to the 1$^{st}$ Tx as $\gamma_0 \to \infty$, and $\alpha_1^{opt}$ is quite close to $m$ for finite but large $\gamma_0$. Referring to (13), this is explained by the fact that 1$^{st}$ step has lowest diversity order ($n$-$m$+1) and hence its error rate dominates. The power allocation algorithm tries to reduce the BLER by allocating more power to the 1$^{st}$ stream and thus reducing the 1$^{st}$ step BER. It should be noted that while the approximation in (20) is more accurate than that in (22), the latter is simpler and more insightful.

*Example:* the average BLER-based optimum power allocation for the $2\times 2$ V-BLAST,

$$\alpha_1^{opt} \approx 2 - \alpha_2^{opt}, \quad \alpha_2^{opt} \approx \sqrt[3]{6/\gamma_0}, \qquad (23)$$

As expected, optimization reduces the dominating 1st step BER,

$$\overline{P}_{e1}^{opt} \approx 1/(8\gamma_0), \qquad (24)$$

if compared to the unoptimized one, $\overline{P}_{e1} = 1/(4\gamma_0)$ [8]. The low power allocated to the 2nd transmitter, $\alpha_2^{opt}$, however, results in decreased diversity order at the 2nd step,

$$\overline{P}_{e2}^{opt} \approx \frac{3}{\left(4\alpha_2^{opt}\gamma_0\right)^2} \approx \frac{3}{16\sqrt[3]{36}\gamma_0^{4/3}}, \qquad (25)$$

so that the optimized 2nd step BER is higher compared to that of the un-optimized system, $\overline{P}_{e2}^{opt} \gg \overline{P}_{e2} \approx 3/(4\gamma_0)^2$. The 1st step BER is therefore reduced at the price of increased 2nd step BER, but the dominant effect of 1st step BER is still preserved in the V-BLAST with optimum power allocation, since $\overline{P}_{e1}^{opt} \gg \overline{P}_{e2}^{opt}$ at sufficiently high SNR. However, this region is achieved at significantly higher SNR ($\sqrt[3]{\gamma_0} \gg 1$) compared to the un-optimized system ($\gamma_0 \gg 1$). Note also that $\overline{P}_{e2}^{opt}$ exhibits a fractional diversity order. Using (12) and (22), it is straightforward to show that $\overline{P}_{e1}^{opt}$ is the dominant contribution to $\overline{P}_{B}^{opt}$ in the general case as well.

Fig. 1 demonstrates that the approximate solution for the optimum power allocation is quite accurate at moderate to high SNR.

*C. Optimum Power Allocation Using the Average TBER*

Similarly to the BLER-based optimization, the average TBER can be used in (16) as an objective function in the Lagrange multiplier technique to find the optimum power allocation. As it was indicated above, the average TBER is convex in $\alpha$ at high SNR; numerical evidence also suggests that it is convex at low to moderate SNR. Hence, the optimum power allocation is unique for arbitrary SNR. Closed-form analytical solution is not feasible due to the complexity of the problem[11]. Since the problem is convex, efficient numerical algorithms can be employed to solve it [9]. At high SNR mode, an accurate approximate closed-form solution can be obtained using the Newton-Raphson method. As in the case of BLER-based optimization, the 1st step BER has the dominant effect on the overall performance [8], and hence the optimization algorithm will have to reduce the 1st step BER by allocating most of the power to the first transmitter, i.e. $\alpha_1^{opt} \to m$, $\alpha_2^{opt}, \ldots, \alpha_m^{opt} \to 0$ as $\gamma_0 \to \infty$. Therefore, for the V-BLAST with "close to optimum" power allocation at high SNR, $\gamma_i^{eff}$, $i = 2, \ldots, m$, are small, and the probability of error propagation can be approximated as $\bar{P}_{ei\|\mathbf{E}_{i-1}\|} \approx 1/2$ if $|\mathbf{E}_{i-1}|^2 \neq 0$. Using this, (8) can be approximated as

$$\bar{P}_{et}(\boldsymbol{\alpha}) = \frac{1}{2m}\sum_{i=1}^{m}(m-i+2)\bar{P}_{ei} \approx \frac{1}{2m}\sum_{i=1}^{m}\frac{(m-i+2)C_{2i-1}^i}{(4\alpha_i\gamma_0)^{n-m+i}} \qquad (26)$$

---
[11] as in the previous case, the Lagrange equations are the system of non-linear transcendental equations

Since $\bar{P}_{et}(\boldsymbol{\alpha})$ is convex (because each term in the sum is convex), the solution to the optimization problem is the unique global minimum. The solution is similar to that for the BLER-based optimization, so that (22) can be used with $b_i$ given by

$$b_i = \sqrt[n-m+i+1]{\frac{C_{2i-1}^i(n-m+i)(m-i+2)m^{n-m+2}}{(m+1)(n-m+1)}}, \qquad (27)$$

and, as before, $b_1 = m$, which means that almost all the power goes to the 1st Tx at high SNR. The difference in the solution for the BLER and TBER-based optimizations is due to the fact that, in the latter case, the error propagation gives a contribution which does not appear in the former case. Since the probability of error propagation is small [8], this difference in the solutions is not large, as we demonstrate below.

Fig. 1 compares the approximate solutions above with the accurate numerical ones. The approximate solutions are already accurate for intermediate to large SNR, $\gamma_0 \geq 5dB$. Moreover, the BLER and TBER-based optimum power allocations are very close to each other and hence the choice of the optimization criteria does not affect significantly the final result. This is not a surprise as the 1st step error rate is dominant, due to the lowest diversity order, in terms of both the average BLER and TBER and hence most of the total power goes to the 1st transmitter, regardless of the criterion.

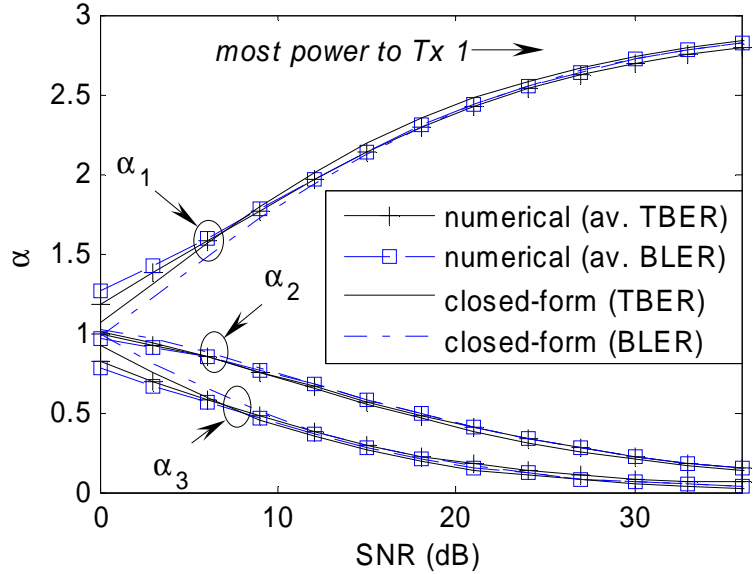

**Fig. 1. Optimum power allocation for 3x3 V-BLAST with BPSK modulation for various optimization strategies.**

*D. Instantaneous versus Average Power Allocation*

In the case of instantaneous optimization, the best power allocation is found for each channel realization. Since an analytical solution is either not feasible or too complicated, with little insight available, a numerical algorithm is used to minimize the instantaneous TBER (see (11), (7), (8)) or BLER (see (2)), subject to total power constraint, for every channel realization **H**. Since $P_{et}$ and $P_B$ are functions of instantaneous per-channel SNRs, the instantaneous feedback of these SNRs to the Tx end is required. As indicated above, the optimum allocation in terms of the BLER is unique for a variety of modulation formats. This uniqueness facilitates numerical evaluation as there is only one global minimum and no local minima. In the case of instantaneous TBER, the solution is not unique so we use several starting points for the numerical algorithm and choose the best solution to insure that the global optimum is found for most of the channel realizations.

The average TBER of instantaneous and average power optimizations is shown on Fig. 2. Clearly, the results are quite close to each other, especially for $\gamma_0 \geq 20 dB$. Essential difference between these two is that the instantaneous optimization performs better in terms of instantaneous TBER, especially for some channel realizations that do not favor the average power allocation. However, the cost of using the instantaneous optimization is higher as each channel instant requires its own optimization and feedback session. On the other hand, the average optimization requires only one computation of $\boldsymbol{\alpha}^{opt}$ as long as the average SNR stays the same. Furthermore, no computationally-expensive numerical optimization is required as the approximate expressions above provide good accuracy, and only $\gamma_0$ needs to be fed back to the transmitter. Thus, the main conclusion here is that the average power optimization can be used instead of instantaneous one at high SNR without any visible penalty in the average error rate, but with much smaller complexity.[12]

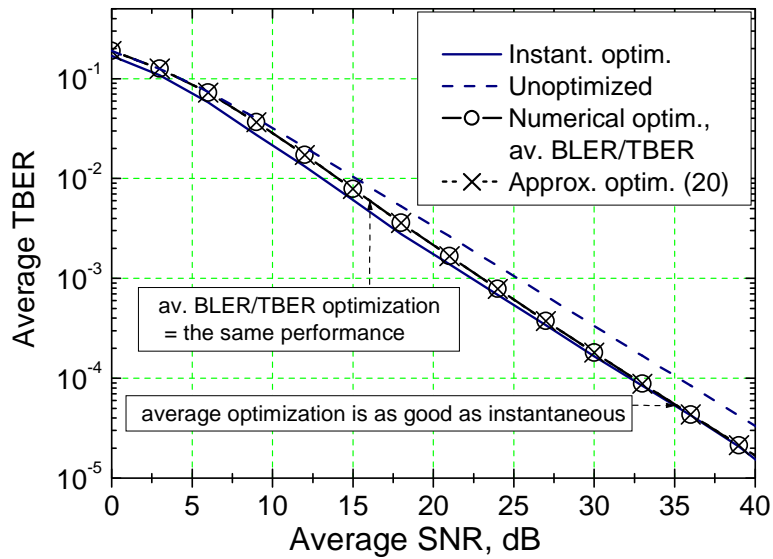

**Fig. 2. Average TBER of 3x3 V-BLAST with BPSK modulation for various optimization strategies.**

IV.  ROBUSTNESS OF THE OPTIMUM POWER ALLOCATION

When the optimization algorithm is implemented in a practical system, there are various sources of inaccuracies and perturbations, which may affect its performance but which were ignored in the idealistic analysis above. These may include numerical inaccuracies of the optimization (due to, for example, limited numerical precisions, i.e. fixed point arithmetic devices, various approximations, etc.), inaccurate or outdated estimate of the average SNR, which result in inaccuracies in the optimum power allocation coefficients $\alpha_i^{opt}$. They may also result from using a finite set of fixed (discrete) power levels instead of continuous power control, as in 2G and 3G systems. A robust algorithm, which is insensitive to all these factors, is desired from the practical perspective.

---

[12] An additional advantage of the average or pre-set (see section IV) power allocation is a simpler design of a power amplifier. Note, however, that all the power allocation algorithms discussed in this paper are compatible with 2G and 3G systems, which employ dynamic power control [20][21]. The proposed algorithms can also accommodate a number of fixed (discrete) power levels, which are used in 2G and 3G systems instead of continuous power allocation, because of algorithms' robustness, as discussed in section IV (see, for example, Fig. 4).

In order to estimate the impact of these factors on the system performance, let us introduce the *measure of robustness* $\delta$ (sensitivity) of the average error rate (either BLER or TBER) with the optimum power allocation, $\overline{P} = \overline{P}(\boldsymbol{\alpha}^{opt})$, to the changes in system parameter $u$,

$$\delta = \left| \frac{\Delta \overline{P} / \overline{P}}{\Delta u / u} \right|, \qquad (28)$$

where $u$ may represent the total Tx power, $u = \sum_{i=1}^{m} \alpha_i$, or the power allocated to any of the transmitters, $u = \alpha_i$. The measure of robustness (28) is the ratio of the normalized variation in the performance $\Delta \overline{P} / \overline{P}$ to the normalized variation in the system parameter $\Delta u / u$, which causes this performance variation. Note that the use of normalized differences in the definition is essential as it makes the measure to be independent of the scale. The algorithm is robust to variations in the system parameter $u$ if relatively small change in $u$ leads to relatively small change in the error rate $\overline{P}$, i.e. when $\delta$ is small or moderate number.

When both the perturbation in the system parameter $\Delta u$ and in the system performance $\Delta \overline{P}$ are small enough, one can use the derivatives in (28) instead of the finite differences,

$$\delta \approx \delta' = \left| \frac{\partial \overline{P}}{\partial u} \frac{u}{\overline{P}} \right|, \qquad (29)$$

so that $\partial \overline{P} / \partial u$ determines the algorithm robustness, and $\delta'$ serves as a measure of local robustness. It follows from (17) that $\partial \overline{P} / \partial \alpha_i = \partial \overline{P} / \partial u = -\lambda$, so that

$$\delta \approx \delta' = \frac{\lambda u}{\overline{P}}, \qquad (30)$$

Thus, the Lagrange multiplier $\lambda$, evaluated at the optimum point and appropriately normalized, is the measure of local sensitivity[13] of the average error rate to variations in the total or individual Tx power. The normalized variation in the average error rate can be evaluated from the normalized variation in the system parameter using (30),

$$\frac{|\Delta \overline{P}|}{\overline{P}} = \delta \frac{|\Delta u|}{u} \approx \delta' \frac{|\Delta u|}{u}, \qquad (31)$$

For the average BLER-based optimization at high SNR, the Lagrange multiplier can be approximated as

---

[13] an extended discussion of this issue in the general framework can be found in [9]

(see Appendix A),

$$\lambda \approx \frac{n-m+1}{m^{n-m+2}} \frac{1}{(4\gamma_0)^{n-m+1}}, \qquad (32)$$

and, from (12), (13), (22), the average optimized BLER is,

$$\overline{P}_B(\mathbf{\alpha}^{opt}) \approx \frac{1}{(4m\gamma_0)^{n-m+1}} \qquad (33)$$

For small variations in the system parameter, $u \approx m \approx \alpha_1$, so that the robustness measure with respect to the variations in the total or 1$^{st}$ transmitter power is

$$\delta_1' \approx n-m+1, \qquad (34)$$

i.e. equal to the diversity order of the system. The algorithm is locally robust as long as $(n-m)$ is not too large; $\delta_1' \approx 1$ and consequently $|\Delta \overline{P}|/\overline{P} \approx |\Delta u|/u$ if $n=m$. This result is a consequence of the fact that the high-SNR average BLER is dominated by the 1$^{st}$ step BER (see (33)) so that its diversity order and hence the sensitivity to the Tx power is minimum when $n=m$; increasing $(n-m)$ results in increasing diversity order and hence in increasing sensitivity to the Tx power. Thus, the beneficial effect of higher diversity order with more Rx antennas is accompanied by the negative effect of higher sensitivity to variations in system parameters.

The robustness measure with respect to $\alpha_2...\alpha_m$ can be approximated as

$$\delta_i' \approx \frac{n-m+1}{m} \frac{b_i}{(4\gamma_0)^{\frac{i-1}{n-m+1+i}}} \ll 1, \quad i=2...m, \qquad (35)$$

where $b_i$ is given by (21). Thus, the algorithm is also robust in terms of $\alpha_2,...,\alpha_m$ at high SNR. Furthermore, higher steps exhibit better robustness since, comparing (34) and (35), $\delta_1' > \delta_i'$, $i=2...m$. It should be noted that this robustness of the algorithm is an unexpected by-product, which was not a goal of the original design.

For the TBER-based optimization, (34) and (35), and hence the conclusions above also hold true; $b_i$ is given by (27) in this case. As an example, Fig. 3 shows the average TBER versus $\alpha_1$ for 2x2 V-BLAST. When $\alpha_1$ is far away from $\alpha_1^{opt}$, the slope of the curves is quite steep and determined by the diversity order of the dominating step; thus, allocating too little power to the 1$^{st}$ Tx increases the 1$^{st}$ step BER, making it dominant, whereas giving too much power to the 1$^{st}$ Tx boosts the 2$^{nd}$ step BER. Note that the slope is steeper in the domain of the dominating 2$^{nd}$ step BER, apparently because of its higher diversity order. But as the power allocation algorithm attempts to balance these two extremes and approaches $\alpha_1^{opt}$, the curves become very flat, confirming local (in the vicinity of $\alpha_1^{opt}$) insensitivity of the TBER to variations in $\alpha_1$.

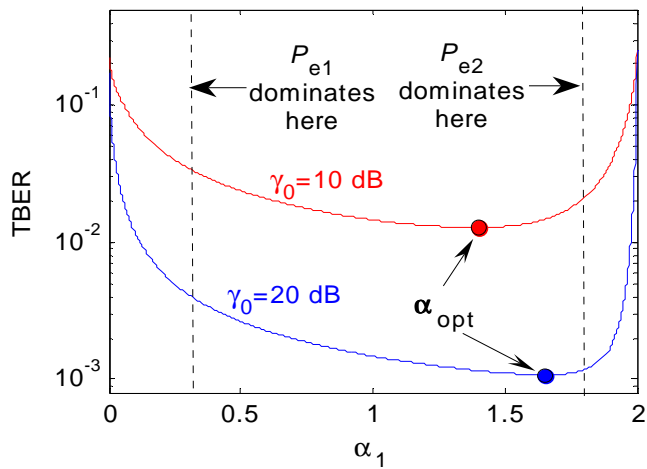

Fig. 3. Average TBER versus $\alpha_1$ for 2x2 V-BLAST with BPSK modulation.

Thus, small inaccuracies in $\boldsymbol{\alpha}^{opt}$ do not affect the average error rate significantly. This hints at the conclusion that the approximate closed-form $\boldsymbol{\alpha}^{opt}$ will result in almost the same average error rate as the accurate numerical one. Fig. 2 confirms this expectation: there is no performance gap between these two. Therefore, the closed-form approximate solution can be used instead of numerical optimization algorithm for the whole range of $\gamma_0$, and not only for $\gamma_0 > 5dB$, where the closed-form solution is more accurate (see Fig. 1). The choice of the optimization criteria (BLER or TBER) does not affect significantly the final result either (since they are indistinguishable on Fig. 2, a single curve represents both). Thus, BLER or TBER can be used equally well as a performance criterion for optimization. From the analytical viewpoint, the BLER (either instantaneous or average) is more advantageous choice since it is convex and has a simple closed-form. It should also be remarked that the effect of the error propagation, which is what differentiates the TBER from the BLER, does not have any significant impact on the optimum power allocation, since both the TBER and BLER are dominated by the 1st step error rate [8].

Small sensitivity of the BLER/TBER to $\boldsymbol{\alpha}$ suggests even further simplification in the optimization algorithm: since $\boldsymbol{\alpha}^{opt}$ changes slowly with the SNR (see Fig. 1), we can pick up only one fixed (pre-set) value of $\boldsymbol{\alpha}$ and still get performance improvement for a wide range of $\gamma_0$. Such simplified algorithm does not require any feedback at all, and yet, as Fig. 4 demonstrates, it attains almost the same performance as the dynamically optimized system. In this example, the 3x3 V-BLAST with $\boldsymbol{\alpha} = [2 \ 0.6 \ 0.4]^T$ is considered, and its performance is very close to the optimized V-BLAST in the range of $\gamma_0 = 0\ldots35$ dB. It follows that the proposed power allocation algorithms can also accommodate a number of fixed power levels, as in 2G and 3G systems [20][21], by quantizing the continuous power allocation (i.e. each SNR interval gets its own pre-set $\boldsymbol{\alpha}$).

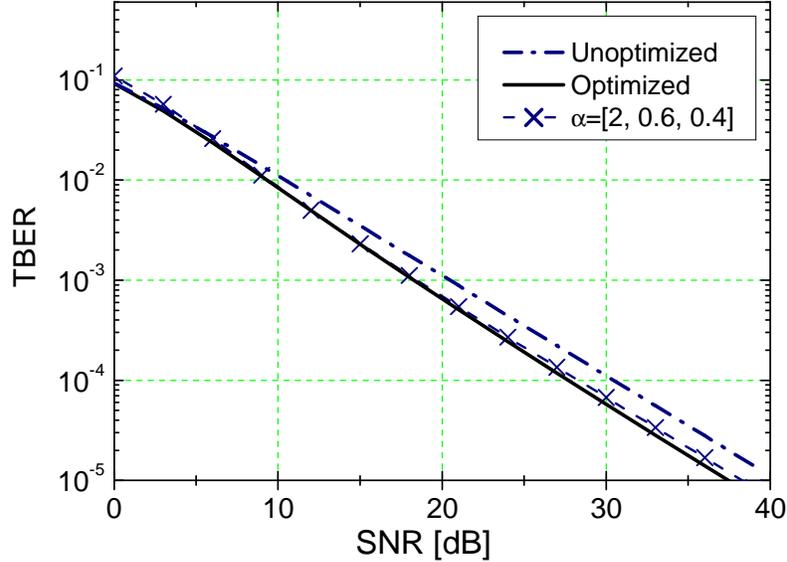

**Fig. 4. TBER of 3x3 V-BLAST with BPSK modulation for pre-set (fixed) power allocation.**

It should also be noted that it is the robustness of the algorithm that is responsible for small difference between instantaneous and average optimization at high SNR.

The robustness considered above is *local* robustness, i.e. for small variations in the vicinity of the unperturbed values of the system parameter. When variations are not small, the finite differences in (28) cannot be accurately approximated by the derivatives and the approximations in (29), (30), (34), (35) may not be accurate. In such a case, one has to consider a measure of *global* robustness. To this end, let us consider the average error rate of the perturbed system $\overline{P}(\boldsymbol{\alpha}; u + \Delta u)$, where $\Delta u$ is not necessarily small and $u$ is the total Tx power. Let $\boldsymbol{\alpha}_{\Delta u}^{opt}$ denote the optimum power allocation of the perturbed system, so that the optimum allocation for the unperturbed system is $\boldsymbol{\alpha}_0^{opt} = \boldsymbol{\alpha}_{\Delta u=0}^{opt}$, the optimized average error rate of the unperturbed system is $\overline{P}(\boldsymbol{\alpha}_0^{opt}; u)$, and the optimized average error rate of the perturbed system is $\overline{P}(\boldsymbol{\alpha}_{\Delta u}^{opt}; u + \Delta u)$. From the general theory of convex optimization [9], the last two quantities are related by the following global inequality,

$$\Delta P = \overline{P}(\boldsymbol{\alpha}_{\Delta u}^{opt}; u + \Delta u) - \overline{P}(\boldsymbol{\alpha}_0^{opt}; u) \geq -\lambda \Delta u,\qquad(36)$$

where $\lambda$ is evaluated at $\Delta u = 0$, and the equality is achieved for $\Delta u = 0$. It follows that if $\Delta u$ is positive, i.e. the total Tx power is increased, the optimal value of $\bar{P}$ decreases by *no more* than $\lambda \Delta u$; if $\Delta u$ is negative, i.e. total Tx power is decreased, the optimal value of $\bar{P}$ is guaranteed to increase by *at least* $\lambda |\Delta u|$. Dividing (36) by $\bar{P}$, one obtains

$$\frac{\Delta \bar{P}}{\bar{P}} \geq -\delta' \frac{\Delta u}{u}, \tag{37}$$

where $\delta'$ is given by (30). Therefore, $\delta'$, which was introduced as a measure of *local* robustness in (30), also serves as a measure of *global* robustness in (37). Since $\Delta \bar{P}, \Delta u$ may be positive as well as negative (they always have opposite sign), we re-write (37) in the form which includes only positive terms,

$$\frac{|\Delta \bar{P}/\bar{P}|}{\Delta u/u} \leq \delta', \ \Delta u > 0; \quad \frac{\Delta \bar{P}/\bar{P}}{|\Delta u/u|} \geq \delta', \ \Delta u < 0 \tag{38}$$

$\delta'$ gives upper and lower bounds on the normalized variation in the error rate due to any (not necessarily small) variation $\Delta u$ in the total Tx power, for positive and negative $\Delta u$ (i.e. increasing and decreasing the total Tx power), respectively. Thus, when the total Tx power is increased by $\Delta u$, the average error rate decreases by not more than $\delta' \bar{P} \Delta u / u$; when the total Tx power is decreased by $\Delta u$, the average error rate increases by at least $\delta' \bar{P} |\Delta u|/u$, so that the positive effect never exceeds the negative one. Since the inequality in (37) transforms into the approximate equality for small perturbations (see (31)), these two effects are equal in that case.

Clearly, the Lagrange multiplier $\lambda$ plays a key role not only in the local, but also in the global robustness. Since it is not known in closed-form for arbitrary SNR, the high-SNR approximation in (32) can be used with reasonable accuracy (for $\gamma_0 > 0$ dB), as Fig. 5 shows.

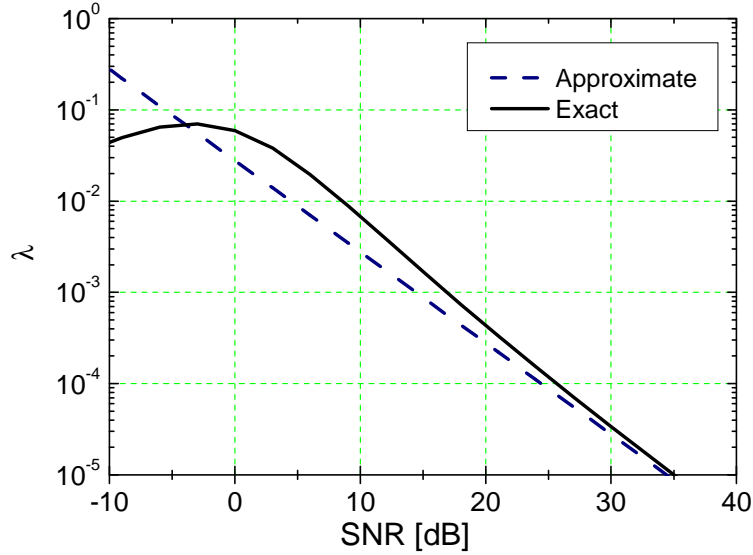

**Fig. 5. The Lagrange multiplier $\lambda$ for the BLER-based optimization versus SNR for 3x3 V-BLAST.**

## V. SNR GAIN OF OPTIMUM POWER ALLOCATION

In this section, we explore some properties of the SNR gain of optimization using mostly analytical techniques, and use numerical results only as the last resort. The analysis and conclusions below are valid for any modulation format, unless otherwise stated.

The *SNR gain* of optimum power allocation is defined as the difference in the SNR required to achieve the same error rate in the optimized and unoptimized systems, i.e. from

$$P\left(\alpha_1^{opt},...,\alpha_m^{opt}\right) = P\left(\alpha,...,\alpha\right), \tag{39}$$

where $P$ is the performance criterion, i.e. the BLER or the TBER, either instantaneous or average; the left-hand side represents the optimized error rate under the total power constraint $\sum_{i=1}^{m}\alpha_i^{opt} = m$, the right-hand side represents the error rate for the uniform power allocation, $\alpha_i = \alpha$, and the SNR gain is $G = \alpha$. For the average optimization, the average error rate is used in (39). For the instantaneous optimization, one may use both instantaneous and average error rate in (39). In the former case, one obtains the instantaneous gain of the instantaneous optimization, and in the latter case, one obtains the average gain (i.e. in terms of the average error rate) of the instantaneous optimization. To be able to compare the instantaneous and average optimizations below, we use the average gain of the instantaneous optimization $G_{inst}$. It compares to the gain of the average optimization $G_{av}$ as follows

$$G_{inst} \geq G_{av} \tag{40}$$

i.e. the instantaneous optimization is at least as good as the average one. This follows from the inequality $\overline{\min_x \{f(x,y)\}} \leq \min_x \{\overline{f(x,y)}\}$, where the expectation is over $y$.

We consider below the SNR gain defined in terms of the BLER and the TBER, and also compare the properties of these two different definitions, which share many similarities.

### A. BLER SNR Gain of the Optimum Power Allocation

In this section, we consider the BLER-based optimization strategies and present universal bounds on the BLER SNR gain, either instantaneous or average, which hold for arbitrary modulation and fading. These results are further refined in the case of BPSK modulation and Rayleigh-fading channel.

*Theorem 1*. The BLER SNR gain of optimum power allocation, either instantaneous or average, for arbitrary modulation and fading, is bounded as follows

$$1 \leq G \leq m \tag{41}$$

*Proof.* The key to the proof is the fact that the BLER, either instantaneous or average, is a monotonically decreasing function in each argument $\alpha_1, ..., \alpha_m$, which follows from (2), (3), and the fact that $P_{ei}$ is a monotonically decreasing function of the SNR. Based on this fact and also on the inequality $P_B(\alpha_1^{opt}, ..., \alpha_m^{opt}) \leq P_B(1, ..., 1)$, which simply states that the optimized system is at least as good as the unoptimized one, the lower bound in (41) follows. Using the monotonic-decreasing property of the BLER and the fact that $\alpha_i \leq m$, which follows from the total power constraint, one concludes that $P_B(\alpha_1^{opt}, ..., \alpha_m^{opt}) \geq P_B(m, ..., m)$. Comparing this inequality with the definition of the gain in (39) in view of the monotonic-decreasing property of the BLER, the upper bound follows. *Q.E.D.*

Below we explore the small-SNR behavior of the SNR gain in terms of the average BLER, which is related to the lower bound in Theorem 1, for the BLER-based optimization and for a variety of modulation formats.

*Theorem 2*. Small-SNR behavior of the BLER SNR gain for the average BLER-based optimization is as follows:

$$G_{av} \to G_0 \geq 1 \text{ as } \gamma_0 \to 0, \tag{42}$$

where

$$G_0 = \begin{cases} \dfrac{m\sum_i a_i^2}{\left(\sum_i a_i\right)^2}, \ a_i = \left.\dfrac{\partial \overline{P}_{ei}}{\partial \sqrt{\alpha_i \gamma_0}}\right|_{\alpha_i \gamma_0 = 0}, & \text{coherent detection} \\ m\dfrac{\max_i |b_i|}{\sum_i |b_i|}, \ b_i = \left.\dfrac{\partial \overline{P}_{ei}}{\partial (\alpha_i \gamma_0)}\right|_{\alpha_i \gamma_0 = 0}, & \text{noncoherent detection} \end{cases} \quad (43)$$

and $a_i$, $b_i$ are the coefficients at the 1$^{st}$ term of MacLaurin's series expansion of $\overline{P}_{ei}$. The equality in (42) is achieved, i.e. $G_0 = 1$, if and only if all $a_i$ or all $b_i$ are equal, for coherent and non-coherent detection respectively.

*Proof*: see Appendix B.

This result is valid for a variety of modulation formats for which the error rate admits MacLaurin's series expansion in SNR or $\sqrt{\text{SNR}}$ about SNR=0. In most cases, the strict inequality in (42) holds, i.e. there is an SNR gain of optimization even at very low SNR, since different $\overline{P}_{ei}$ exhibit different behavior so that the expansion coefficients are also different. For coherent BPSK and non-coherent BFSK,

$$a_i^{BPSK} = -\frac{n-m+i}{2} + \frac{1}{2^{n-m+i}} \sum_{k=0}^{n-m+i-1} C_{n-m+i+k-1}^k \frac{k}{2^k}, \ b_i^{BFSK} = a_i^{BPSK}/2, \quad (44)$$

and $G_0 = 0.17$ dB and $0.79$ dB respectively for 2x2 system.

*Corollary 2.1.* The result in (42), (43) also applies to the instantaneous gain of the instantaneous optimization, in which case $P_{ei}$ should be used in (43) instead of $\overline{P}_{ei}$, and the coefficients $a_i$, $b_i$ and hence the gain depend on the channel realization, as long as the derivatives in (43) exist and are not all equal to zero simultaneously.

*Corollary 2.2.* The average SNR gain of the instantaneous optimization is also lower bounded by $G_0$ in (43), $G_{inst} \geq G_0 \geq 1$, because of (40).

Thus, we conclude that (42) holds for a variety of scenarios for BLER-based optimization.

We now show that the upper bound in (41) is achieved at high SNR.

*Theorem 3.* High-SNR behavior of the average BLER SNR gain, for both instantaneous and average optimizations using either BLER or TBER as an objective, with BPSK modulation in Rayleigh fading channel, is as follows:

$$G \to m \text{ as } \gamma_0 \to \infty \quad (45)$$

*Proof.* From the high SNR approximation of the average BLER (12) with $\boldsymbol{\alpha}^{opt}$ given by (22), one

Accepted by IEEE Trans. Comm. 23(35)

concludes that

$$\overline{P}_B \to \overline{P}_{e1} = \begin{cases} \overline{P}^{MRC}_{n-m+1}(\gamma_0), & \text{unoptimized} \\ \overline{P}^{MRC}_{n-m+1}(m\gamma_0), & \text{optimized} \end{cases} \quad \text{as } \gamma_0 \to \infty \qquad (46)$$

i.e. the first step dominates for both unoptimized and optimized systems. Using this in the gain definition in (39), one concludes that for the average optimization, $G_{av} \to m$ as $\gamma_0 \to \infty$. Using (41) and (40), this also holds for the average gain of the instantaneous optimization, $G_{inst} \to m$ as $\gamma_0 \to \infty$. Q.E.D.

*Corollary 3.1.* Theorem 3 also extends to any modulation/fading for which (46) holds, i.e. for which the first step error rate dominates the average BLER at high SNR. Based on the diversity order argument, this condition should hold for most modulation formats in Rayleigh-fading channels.

For the average BLER-based optimization with BPSK modulation in a Rayleigh fading channel, a high-SNR approximation of the average BLER SNR gain is given by

$$G_{av} \approx \frac{G_\infty}{\sqrt[n-m+1]{1 + c_{m,n}/\sqrt[n-m+3]{4\gamma_0}}}, \quad c_{m,n} = \frac{(n-m+1)b_2^{n-m+3} + 3m^{n-m+2}}{mb_2^{n-m+2}} \qquad (47)$$

where $G_\infty = m$ and $b_2$ is given by (21). This approximation follows along the lines of the proof of Theorem 3: first, a high SNR approximations of the average BLER of the optimized and unoptimized systems are obtained by keeping the first two terms of the Taylor series expansion of the exact BLER expressions about $1/\gamma_0 = 0$; secondly, they are used in the gain definition in (39) to obtain (47). Note that (47) reduces to (45) for $\gamma_0 \to \infty$, as it should be. The convergence to the upper bound in (45) is however slow, since the convergence condition is $\sqrt[n-m+3]{\gamma_0} \gg 1$.

It follows from (47) that the BLER SNR gain of the average BLER-based optimization with BPSK modulation is an increasing function of the average SNR in the high-SNR range. Numerical evidence indicates that this also holds for low to intermediate SNR, see Fig. 6. This conclusion is further reinforced by the following Theorem.

*Theorem 4.* Under the total power constraint $\sum_{i=1}^{m} \alpha_i = u$, where $u$ is the total Tx power, the BLER SNR gain of BLER-based optimization in (39), either instantaneous or average, is a monotonically increasing function of $u$:

$$\frac{\partial G}{\partial u} = -\frac{\lambda}{\partial P_B(\alpha\ldots\alpha)/\partial \alpha} \geq 0 \qquad (48)$$

*Proof.* see Appendix B.

### B. TBER SNR Gain of the Optimum Power Allocation

In this section, we adapt the results of the previous section to the SNR gain defined in terms of the average TBER. To accomplish this, we will need the following generic properties of the average TBER, which serve as a substitution to the monotonically decreasing property of the BLER[14].

*Property 1.* For regular (unoptimized) V-BLAST, the average TBER decreases with SNR or total Tx power,

$$\partial \bar{P}_{et}(\alpha,\ldots,\alpha)/\partial \alpha < 0 \qquad (49)$$

This property is supported by the results in [6], [8], and also by extensive results in the literature on the error rate performance of diversity systems with co-channel interference [10]-[13][15], since the error propagation in V-BLAST effectively "creates" co-channel interference and thus it is similar to diversity combining with interference. Note that since all the Tx powers are increased in (49) simultaneously, the effective signal-to-interference ratio for propagating errors stays the same or even decreases due to decrease in the probability of error propagation [8].

*Property 2.* For a partially-optimized V-BLAST, in which $\alpha_i^{opt}$ are used up to the stream $k<m$, the TBER decreases when the SNR at the unoptimized streams increases:

$$\partial \bar{P}_{et}\left(\alpha_1^{opt},\ldots,\alpha_k^{opt},\alpha\ldots,\alpha\right)/\partial \alpha < 0 \qquad (50)$$

This property follows from Property 1 and (5), since first $k$ errors rates $P_{u1},\ldots,P_{uk}$ are fixed and the others are those of the unoptimized system and behave according to Property 1, i.e. decrease with the SNR. The same reasoning also holds for instantaneous TBER.

*Property 3.* For the average TBER-based optimization, the following inequality holds,

$$\alpha_{k-1}^{opt} \geq \alpha_k^{opt}, \ k = 2\ldots m \qquad (51)$$

---

[14] Unfortunately, the TBER is not necessarily monotonically decreasing in $\alpha_1\ldots\alpha_m$ since increasing power of some transmitters at lower steps increases inter-stream interferences to the others at higher steps due to the effect of error propagation. Hence, the TBER may potentially increase if the error propagation effect is strong enough.

[15] to the best of our knowledge, it has never been observed in the literature that error rate may increase with SNR, even when co-channel interference is present. This strongly supports Property 1.

The rationale behind this property is the fact that according to the optimum power allocation strategy, more power is granted to the more significant source of errors. Since lower steps generate on average more errors due to lower average after-processing SNR[16] and the effect of error propagation, they are allocated more power. This property is also supported by numerical evidence. This, however, does not hold for the instantaneous TBER-based optimization since some channel realization may "favor" lower steps, which may thus have higher SNR and fewer errors than higher steps, so that they will get less Tx power.

We can now show that Theorem 1 also holds for the average TBER SNR gain. The lower bound in (41) follows from the fact that optimization can not increase the TBER, so that $\bar{P}_{et}(\boldsymbol{\alpha}^{opt}) \geq \bar{P}_{et}(1,\ldots,1)$; using Property 1 gives the lower bound.

To prove the upper bound in (41) we note that by Property 2 for $k = m-1$, $\bar{P}_{et}(\alpha_1^{opt},\ldots,\alpha_{m-1}^{opt},\alpha)$ is a decreasing function of $\alpha$, and by Property 3, $\alpha_{m-1}^{opt} \geq \alpha_m^{opt}$; therefore, $\bar{P}_{et}(\alpha_1^{opt},\ldots,\alpha_{m-1}^{opt},\alpha_m^{opt}) \geq \bar{P}_{et}(\alpha_1^{opt},\ldots,\alpha_{m-1}^{opt},\alpha_{m-1}^{opt})$. Repeating this for $k = m-2,\ldots,1$, one obtains

$$\bar{P}_{et}(\boldsymbol{\alpha}^{opt}) \geq \ldots \geq \bar{P}_{et}(\alpha_1^{opt},\ldots,\alpha_1^{opt}) \geq \bar{P}_{et}(m,\ldots,m) \quad (52)$$

The last inequality here is due to Property 1 and $\alpha_i \leq m$, which follows from the total power constraint. Using (52) in the gain definition in (39) and relying on Property 1, the upper bound follows.

For the SNR gain defined in terms of instantaneous TBER for instantaneous optimization, analytical proof of the upper-bound in (41) presents serious difficulties, but numerical evidence suggests that it is still valid.

Theorem 2 still holds for the average TBER SNR gain, with the substitution of $\bar{P}_{ei} \to \bar{P}_{et}$ in (43).

Theorem 3 is no longer valid, i.e. the upper bound $m$ is never attained if the gain is defined in terms of the TBER. Instead, the following holds.

*Theorem 5.* High SNR behavior of the average TBER SNR gain for the average optimization is as follows:

$$G_{av} \to G_\infty \text{ as } \gamma_0 \to \infty, \quad G_\infty = m\left(\frac{2\bar{a}_1}{m+1}\right)^{\frac{1}{n-m+1}} < m, \quad (53)$$

where $\bar{a}_1$ is given by (14).

*Proof.* From the high SNR approximations of the average TBER for optimized and unoptimized systems

---

[16] projecting out interference from yet-to-be-detected higher-step symbols results in SNR loss. Since the dimensionality of the projected out space decreases for higher steps, the SNR loss decreases as well.

in (14) and (26), one concludes that

$$\overline{P}_{et} \to \begin{cases} \dfrac{\overline{a}_1}{m} \dfrac{1}{(4\gamma_0)^{n-m+1}}, & \text{unoptimized} \\ \dfrac{m+1}{2m} \dfrac{1}{(4m\gamma_0)^{n-m+1}}, & \text{optimized} \end{cases}, \text{ as } \gamma_0 \to \infty \quad (54)$$

Using (54) in (39), it follows that $G_{av} \to G_\infty$ as $\gamma_0 \to \infty$. Q.E.D.

Thus, the improvement in average TBER is less than the upper-bound in (41). The reason for this is the increased power of propagating errors for the optimized system, due to higher power going to lower steps, compared to the unoptimized one. For example, the optimum power allocation algorithm gives most of the power to the 1$^{st}$ Tx trying to avoid the errors at the 1$^{st}$ step. But if the error *does* occur at the 1$^{st}$ step, its amplitude is higher than that for the unoptimized system, which makes the error propagation effect more severe.

The high-SNR approximation in (47) holds for the TBER SNR gain, with $G_\infty$ given by (53) and

$$c_{m,n} = \frac{(m+1)(n-m+1)b_2^{n-m+3} + 3m^{n-m+3}}{m(m+1)b_2^{n-m+2}}, \quad (55)$$

where $b_2$ is given by (27).

### C. Examples and Comparisons

Below we consider the $2 \times 2$ V-BLAST to get some insight. The average BLER at high SNR is,

$$\overline{P}_B \approx \begin{cases} \overline{P}_{e1}(\gamma_0) \approx 1/(4\gamma_0), & \text{unoptimized} \\ \overline{P}_{e1}(2\gamma_0) \approx 1/(8\gamma_0), & \text{optimized} \end{cases} \quad (56)$$

so that $G_{av} = 2$ (3 dB), which is the same as upper-bound in (41). This is not the case for the average TBER SNR. At high SNR, the probability of error propagation for the unoptimized V-BLAST is $\overline{P}_{e2|2} \approx 1/5$ [8], and $\overline{P}_{e2|2} \approx 1/2$ for the optimized V-BLAST. The average TBER is,

$$\overline{P}_{et} \approx \overline{P}_{e1}(1 + \overline{P}_{e2|2})/2 \approx \begin{cases} 3\overline{P}_{e1}(\gamma_0)/5 \approx 3/(20\gamma_0), & \text{unoptimized} \\ 3\overline{P}_{e1}(2\gamma_0)/4 \approx 3/(32\gamma_0), & \text{optimized} \end{cases} \quad (57)$$

so that $G_{av} = 8/5$ (2 dB) at high SNR, which is less than the 3 dB upper-bound in (41). The high-SNR behavior of the average BLER and TBER gains is as follows,

$$G_{av,BLER} \approx 2\left[1+\frac{9}{2\sqrt[3]{36\gamma_0}}\right]^{-1}, \quad G_{av,TBER} \approx \frac{8}{5}\left[1+\frac{3}{2\sqrt[3]{2\gamma_0}}\right]^{-1} \tag{58}$$

The condition for convergence to the upper bound in (58) is $\sqrt[3]{\gamma_0} \gg 1$, i.e. the convergence is slow. Specifically, the upper bound is achieved approximately when the second term in the brackets in (58) does not exceed 0.1, which results in $\gamma_0 \geq 35 dB$ and $\gamma_0 \geq 30 dB$ for the BLER and TBER gains respectively.

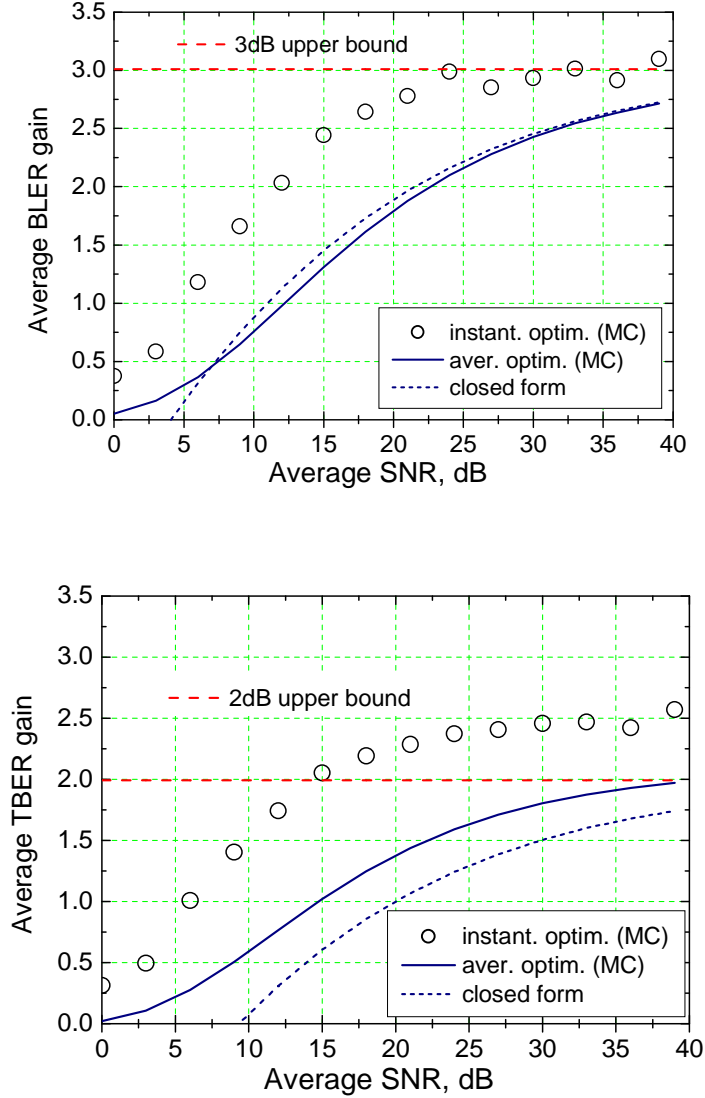

**Fig. 6. Average BLER (a) and TBER (b) SNR gains vs. SNR for 2x2 V-BLAST with BPSK modulation (MC- Monte-Carlo simulations).**

The simulation results validate these conclusions. Fig. 6(a) demonstrates that the average BLER gain of the optimized $2\times 2$ V-BLAST monotonically increases with SNR and tends to 3 dB, which is the upper bound. The average BLER gain of the instantaneous optimization also tends to 3 dB and attains this bound, but much faster than that of the average optimization. The average TBER SNR gain of the same system approaches 2 dB at high SNR, monotonically and slowly increasing. For the instantaneous optimization, this gain is higher and it approaches higher than 2 dB upper bound faster. Thus, in the case of TBER-defined gain, an additional advantage of using the instantaneous optimization is that its average SNR gain achieves higher upper bound than that of the average optimization at high SNR.

Whether the gain is defined using the TBER or the BLER, it exhibits roughly the same behavior: it is bounded as in (41), it approaches the lower bound at low SNR, and it increases monotonically with the SNR. In both cases, the instantaneous optimization attains the upper-bound at $\gamma_0 \approx 20...25$ dB. In both cases, the gain of average optimization increases with the SNR much slower than that of the instantaneous one. The only significant difference is that the BLER gain of average optimization attains the upper bound = m, and the TBER gain does not. However, since the upper bound is achieved by the BLER SNR gain at rather high SNR, this difference seems to be less important from practical perspective. As indicated by Fig. 6(a) and 6(b), both gains are close to each other at a practical SNR range.

The SNR gain of the optimum power allocation is almost the same, at high SNR, as that of the optimal ordering procedure (see [7] for details). The computational complexity, however, of the former is much less than that of the latter. Hence, the average power optimization can be used instead of the optimal ordering with roughly the same performance.

## VI. CONCLUSION

Based on recent results on the error rate performance of the unordered V-BLAST [6], [8], optimum power allocation techniques for this algorithm, which are based on various criteria, i.e. average and instantaneous block and total error rates, have been systematically studied and compared to each other. Accurate high-SNR approximations in a compact closed form have been derived for the optimum power allocation based on average BLER and TBER. Due to the dominant contribution of lower steps to the error rate, these steps get more power in the optimized system, with the 1$^{st}$ step getting most of it. The allocation strategies based on the average BLER and TBER are almost the same, since in both cases 1$^{st}$ step is dominant in terms of the average error rate.

The error rate performance of the optimized system has been investigated; closed-form expressions for optimized error rates at high SNR have been obtained. In spite of the non-uniform power allocation in the optimized system, 1$^{st}$ step still dominates in terms of the error rate. Conditional error rates of the higher steps exhibit fractional diversity order. While the instantaneous optimization requires instantaneous feedback and thus is more complex than the average one, which requires only the average SNR be fed back to the transmitter, its average error rate performance is only slightly better than that of the average optimization. At high SNR, both strategies exhibit the same average error rate. The average optimum power allocation can be considered as a low-complexity alternative to the optimal ordering.

The SNR gain of the optimization has been rigorously defined for various criteria (average and instantaneous BLER and TBER) and investigated. Universal bounds have been derived, which hold true for any optimization strategy, modulation format and channel fading type. Specifically, the gain is upper bounded by the number of transmit antennas. The upper bound is achieved at high SNR for the average BLER-defined SNR gain, but is not for the TBER-defined gain, which is due to increased power of the error propagation in the optimized system. Closed-form high SNR approximations of the gain have been obtained.

Robustness of the optimized algorithm in terms of variations in total and individual Tx powers have been studied. Using the measure of robustness introduced, it has been demonstrated that the optimized algorithm is not sensitive to small variations in the above-mentioned parameters. This suggests a simplified power allocation strategy, which does not require any feedback at all but rather relies on pre-set values of transmit powers. The performance of this strategy is close to that of dynamical optimization over the practical SNR range.

The analytical results and approximations above have been validated via simulations.

## VII. ACKNOWLEDGEMENT

The authors would like to thank F. Gagnon for numerous stimulating discussions.

Systems, 2005 IEEE International Conference on Communications, San Diego, California USA, v. 4, pp. 2282 – 2286, May 2005.

[5] N. Wang, S.D. Blostein, Minimum BER Transmit Power Allocation and Beamforming for Two-Input Multiple-Output Spatial Multiplexing Systems, 2005 Canadian Workshop on Information Theory, Montreal, Canada, Jun. 2005.

[6] N. Prasad, M.K. Varanasi, Analysis of decision feedback detection for MIMO Rayleigh-fading channels and the optimization of power and rate allocations, IEEE Trans. Inform. Theory, v. 50, No. 6, June 2004

[7] S. Loyka, F. Gagnon, Performance Analysis of the V-BLAST Algorithm: an Analytical Approach, IEEE Trans. Wireless Comm., v. 3, No. 4, pp. 1326-1337, July 2004.

[8] S. Loyka, F. Gagnon, V-BLAST without Optimal Ordering: Analytical Performance Evaluation for Rayleigh Fading Channels, IEEE Transactions on Communications, v. 54, N. 6, pp. 1109-1120, June 2006.

[9] S. Boyd, L. Vandenberghe, Convex Optimization, Cambridge University Press, 2004.

[10] M. K. Simon, M.-S. Alouini, Digital Communication over Fading Channels, Wiley, 2005.

[11] A. Shah, A.M. Haimovich, Performance analysis of maximal ratio combining and comparison with optimum combining for mobile radio communications with cochannel interference, IEEE Trans. Vehicular Technology, v. 49, No 4, pp.1454 – 1463, Jul. 2000

[12] J.S. Kwak, J. H. Lee, Closed-form expressions of approximate error rates for optimum combining with multiple interferers in a Rayleigh fading channel, IEEE Trans. Vehicular Technology, v. 55, No. 1, pp. 158 – 166, Jan. 2006

[13] M. Chiani, M.Z.Win, A. Zanella, Error probability for optimum combining of M-ary PSK signals in the presence of interference and noise, IEEE Trans. Comm., v. 51, No. 11, pp.1949 – 1957, Nov. 2003

[14] G. Ginis, J.M. Cioffi, On the Relation Between V-BLAST and the GDFE, IEEE Communications Letters, v. 5, N. 9, pp. 364-366, Sep. 2001.

[15] S. Loyka, V. Kostina, F. Gagnon, On Convexity/Concavity Properties of Error Rates of the ML Detector and Their Applications, IEEE Trans. Information Theory, submitted, 2006.

[16] S. Loyka, F. Gagnon, Analytical Framework for Outage and BER Analysis of the V-BLAST Algorithm, 2004 IEEE International Zurich Seminar on Communications, Feb. 18-20 2004, ETH Zurich, Switzerland, pp. 120-123.

[17] G.A. Korn, T.M. Korn, Mathematical Handbook for Scientists and Engineers, Dover, New York, 2000.

[18] J.G. Proakis, Digital Communications, McGraw Hill, Boston, 2001.

[19] J. Choi, Nulling and Cancellation Detector for MIMO Channels and its Application to Multistage
Accepted by IEEE Trans. Comm.                                                                                      31(35)

Receiver for Coded Signals: Performance and Optimization, IEEE Trans. Wireless Communications, v.5, N.5, pp. 1207-1216, May 2006.

[20] R. Pichna, Q. Wang, Power Control, in J.D. Gibson (Ed.), Communications Handbook, CRC Press, Boca Raton, 2002, pp. 78-1 – 78.11.

[21] A.F. Molisch, Wireless Communications, IEEE Press and Wiley, 2005.

[22] H. Lee, I. Lee, New Approach to Error Compensation in Coded V-BLAST OFDM Systems, IEEE Trans. on Communications, v. 55, N. 2, pp. 345-355, Feb. 2007.

Appendix A. CLOSED-FORM EXPRESSIONS FOR OPTIMAL POWER ALLOCATION

In the high SNR range, $\bar{P}_B$ is approximated by (12), and (17) is transformed into

$$\frac{\partial L(\boldsymbol{\alpha})}{\partial \alpha_i} = -\frac{(n-m+i)C_{2i-1}^i}{(4\gamma_0)^{n-m+i}\alpha_i^{n-m+i+1}} + \lambda = 0, \quad i=1\ldots m \tag{A1}$$

It follows that each $\alpha_i$ can be expressed via the single parameter $\lambda$:

$$\alpha_i = \frac{a_i}{\sqrt[n-m+i+1]{(4\gamma_0)^{n-m+i}\lambda}}, \quad a_i = \sqrt[n-m+i+1]{(n-m+i)C_{2i-1}^i}, \tag{A2}$$

Substituting (A2) into the total power constraint $\sum_{i=1}^{m}\alpha_i = m$ and introducing a new variable $x = \lambda^{-(n+1)!/(n-m+1)!}$, we obtain the following equation for $x$:

$$\sum_{i=1}^{m}(4\gamma_0)^{-\frac{n-m+i}{n-m+i+1}} a_i x^{c_i} = m, \quad c_i = \frac{(n+1)!}{(n-m+1)!(n-m+i+1)} \tag{A3}$$

Following the Newton-Raphson method [17], the zero-order approximate solution to (A3) is found when all only the leading term ($i=1$) is kept in (A3),

$$x_0 = \sqrt[c_1]{m(4\gamma_0)^{\frac{n-m+1}{n-m+2}}/a_1} \tag{A4}$$

First-order approximate solution can be written as $x = x_0(1+\delta)$, where $\delta$ is a small increment. By substituting $x$ into the left-hand side of (A3), one obtains

$$\sum_{i=1}^{m}a_i x_0^{c_i}(1+\delta)^{c_i} = \sum_{i=1}^{m}b_i(1+\delta)^{c_i}(4\gamma_0)^{\frac{1-i}{n-m+i+1}} = m,$$

$$b_i = a_i a_1^{-\frac{n-m-2}{n-m+i+1}} m^{\frac{c_i}{c_1}} = \sqrt[n-m+i+1]{\frac{(n-m+i)m^{n-m+2}C_{2i-1}^i}{n-m+1}} \tag{A5}$$

Keeping only the first two leading terms in (A5), $i=1,2$, one obtains:

$$m(1+c_1\delta)+b_2(4\gamma_0)^{-1/(n-m+3)}=m, \quad (A6)$$

from which it follows that

$$\delta=-b_2(mc_1)^{-1}(4\gamma_0)^{-1/(n-m+3)} \quad (A7)$$

and

$$x=x_0(1+\delta)=\sqrt[c_1]{a_1^{-1}m(4\gamma_0)^{\frac{n-m+1}{n-m+2}}}\left[1-\frac{b_2}{mc_1\sqrt[n-m+3]{4\gamma_0}}\right],$$

$$\lambda=x^{-(n-m+2)c_1}=\frac{1}{(4\gamma_0)^{n-m+1}}\left(\frac{a_1}{m}\right)^{n-m+2}\left[1-\frac{b_2}{mc_1\sqrt[n-m+3]{4\gamma_0}}\right]^{-(n-m+2)c_1} \quad (A8)$$

Finally, the optimum power allocation $\boldsymbol{\alpha}^{opt}$ is found by substituting $\lambda$ from (A8) into (A2):

$$\tilde{\alpha}_i=b_i(4\gamma_0)^{\frac{1-i}{n-m+i+1}}\left(1-\frac{b_2}{mc_1\sqrt[n-m+3]{4\gamma_0}}\right)^{c_i}, \quad \alpha_i^{opt}=m\tilde{\alpha}_i/\sum_{k=1}^{m}\tilde{\alpha}_k \quad (A9)$$

The last equality here is to assure the total power constraint for the approximate solution.

Appendix B.  THE GAIN OF OPTIMIZATION FOR SMALL SNR

The main idea of the proof of Theorem 2 is to use the small SNR approximations of $\bar{P}_B$ to find $\boldsymbol{\alpha}^{opt}$. The gain is then found in closed form using the definition (39). The small SNR approximation of $\bar{P}_{ei}$ is found via the MacLaurin's series expansion. For coherent modulation formats, the error rate is a function of $\sqrt{SNR}$ [18] so that the expansion in $\sqrt{\alpha_i\gamma_0}$ about zero is used[17] as a low-SNR approximation,

$$\bar{P}_{ei}\approx\frac{1}{2}+a_i\sqrt{\gamma_0}, \quad a_i=\left.\frac{\partial\bar{P}_{ei}}{\partial\sqrt{\alpha_i\gamma_0}}\right|_{\alpha_i\gamma_0=0}. \quad (B1)$$

where we keep only the two leading terms. For the BPSK modulation, $a_i^{BPSK}$ are given by (44). The small SNR approximation of the BLER is obtained by using (B1) in (3) and keeping only first-order terms in $\sqrt{\alpha_i\gamma_0}$,

$$\bar{P}_B(\boldsymbol{\alpha})\approx 1-\frac{1}{2^m}+\frac{1}{2^{m-1}}\sum_{i=1}^{m}a_i\sqrt{\gamma_0\alpha_i}. \quad (B2)$$

Using (B2) as the objective function in (17), one finds $\boldsymbol{\alpha}^{opt}$:

---

[17] recall that $\sqrt{x}$ cannot be expanded in MacLaurin's series.

$$\left.\frac{\partial L(\boldsymbol{\alpha})}{\partial \alpha_i}\right|_{\boldsymbol{\alpha}=\boldsymbol{\alpha}^{opt}} = \frac{a_i}{2^m}\sqrt{\frac{\gamma_0}{\alpha_i^{opt}}} + \lambda = 0 \rightarrow \alpha_i^{opt} = \frac{ma_i^2}{\sum_{i=1}^m a_i^2}, \quad (B3)$$

where we have also used the constraint $\sum_{i=1}^m \alpha_i^2 = m$ to find $\lambda$. The optimized BLER can now be expressed as $\bar{P}_B(\boldsymbol{\alpha}^{opt}) = 1 - 2^{-m} + 2^{-m+1}\sqrt{m\gamma_0}\sqrt{\sum_i a_i^2}$. Comparing it to the non-optimized one in the gain definition in (39), $\bar{P}_B(\alpha,\alpha,...,\alpha) = 1 - 2^{-m} + 2^{-m+1}\sqrt{\alpha\gamma_0}\sum_i a_i$, one finds the gain $G_0 = \alpha$,

$$G_0 = \frac{m\sum_i a_i^2}{\left(\sum_i a_i\right)^2} \geq 1 \quad (B4)$$

By the Cauchy-Schwartz inequality, the equality holds if all $a_i$ are equal. We note that the same argument holds for the instantaneous BLER and hence (B4) also holds as long as the approximation in (B1) applies, which is the case for BPSK modulation, as well as for some other modulation formats [18].

For non-coherent modulation formats (e.g. DPSK, FSK), the MacLaurin's series expansion in terms of SNR can be used as a low-SNR approximation of the error rate,

$$\bar{P}_{ei} \approx \frac{1}{2} + b_i \gamma_0, \ b_i = \left.\frac{\partial \bar{P}_{ei}}{\partial (\alpha_i \gamma_0)}\right|_{\alpha_i \gamma_0 = 0}, \quad (B5)$$

For the BFSK detected non-coherently, $b_i^{BFSK}$ are given by (44). Similarly to the coherent case, the small SNR approximation of the BLER is,

$$\bar{P}_B(\boldsymbol{\alpha}) \approx 1 - \frac{1}{2^m} + \frac{1}{2^{m-1}}\sum_{i=1}^m b_i \alpha_i \gamma_0. \quad (B6)$$

The derivatives $\partial L(\boldsymbol{\alpha})/\partial \alpha_i = -b_i 2^{m-1}\gamma_0 + \lambda$ cannot be now all equal to zero simultaneously unless $b_i$ are the same, which means that there exists no stationary point inside of the feasible region and the solution is located on the region boundary. It follows from (B6) that the BLER is minimized when all the power is allocated to the stream for which $|b_i|$ is maximum[18],

$$\alpha_{i_{\max}}^{opt} = m, \ \alpha_i^{opt} = 0, \ i \neq i_{\max}, \ \text{where } i_{\max} = \arg\max_i |b_i| \quad (B7)$$

When all $b_i$ are equal, any power allocation gives the same BLER. Finally, using (B7) in the definition of the SNR gain (39), one obtains:

---

[18] the magnitude is required since all $b_i$ are non-positive

$$G_0 = m \frac{\max_i |b_i|}{\sum_i |b_i|} \geq 1. \tag{B8}$$

The equality holds if all $b_i$ are equal, in which case, as expected, there is no gain due to the optimization. It should be pointed out that the argument above holds for the instantaneous BLER as long as the approximation in (B5) holds, which is the case for many non-coherent modulation formats [18]. Thus, (B8) also applies to the instantaneous gain of the instantaneous optimization, in which case $b_i$ and hence the gain depend on channel realization.

For the instantaneous BLER-based optimization, the averaged over $\mathbf{H}$ gain cannot be worse than that of the average optimization, see (40), hence the former is also subject to the inequalities in (B8) and (B4).

The proof of Theorem 4 is as follows. From the total power constraint,

$$\sum_{i=1}^{m} \frac{\partial \alpha_i^{opt}}{\partial u} = 1, \tag{B9}$$

and from the optimality condition (16)

$$\partial \bar{P}_B\left(\boldsymbol{\alpha}^{opt}\right) / \partial \alpha_i^{opt} = -\lambda \leq 0 \tag{B10}$$

Taking the derivative of (39) with respect to $\alpha_i^{opt}$,

$$\frac{\partial \bar{P}_B\left(\boldsymbol{\alpha}^{opt}\right)}{\partial \alpha_i^{opt}} = \frac{\partial \bar{P}_B(\alpha\ldots\alpha)}{\partial \alpha} \frac{\partial \alpha}{\partial \alpha_i^{opt}} \rightarrow \frac{\partial \alpha}{\partial \alpha_i^{opt}} = \frac{\partial \bar{P}_B\left(\boldsymbol{\alpha}^{opt}\right)}{\partial \alpha_i^{opt}} \bigg/ \frac{\partial \bar{P}_B(\alpha\ldots\alpha)}{\partial \alpha}, \tag{B11}$$

Using (B9)-(B11), one obtains:

$$\frac{\partial \alpha}{\partial u} = \sum_{i=1}^{m} \frac{\partial \alpha}{\partial \alpha_i^{opt}} \frac{\partial \alpha_i^{opt}}{\partial u} = -\frac{\lambda}{\partial \bar{P}_B(\alpha\ldots\alpha)/\partial \alpha} \geq 0 \tag{B12}$$

The same reasoning holds true if the instantaneous BLER is used to define the SNR gain.